\documentclass{PoS}
\usepackage{xcolor}
\usepackage{amsmath}
\usepackage{slashed}
\usepackage{stmaryrd}
\usepackage{bbm}
\usepackage{braket}
\usepackage{wrapfig}

\definecolor{mygreen}{HTML}{278e25}
\def\Dslash{\slashed{D}}
\def\pbp{\bar\psi\psi}

\title{Exploring the QCD phase diagram via reweighting from isospin chemical potential}

\ShortTitle{QCD phase diagram via reweighting}

\author{B. B. Brandt, F. Cuteri, G. Endr\H{o}di and \speaker{S. Schmalzbauer}\thanks{The research has been funded by the DFG via the Emmy Noether Programme EN 1064/2-1. S.S. has also received support from HGS-HIRe for FAIR and the Stiftung Giersch.}\\
        Institute for Theoretical Physics, Goethe University, Max-von-Laue-Strasse 1, 60438 Frankfurt am  Main, Germany\\
        E-mail: \email{\{brandt,cuteri,endrodi,schmalzbauer\}@itp.uni-frankfurt.de}}

\abstract{We investigate the QCD phase diagram for small values of baryon and strange quark
chemical potentials from simulations at non-zero isospin chemical potential. Simulations
at pure isospin chemical potential are not hindered by the sign problem and pion condensation
can be observed for sufficiently large isospin chemical potentials. We study how the related
phase boundary evolves with baryonic and strange chemical potentials via reweighting in quark
chemical potentials and discuss our results. Furthermore, we propose and implement an
alternative method to approach nonzero baryon (and strange quark) chemical potentials.
This method involves simulations where physical quarks are paired with auxiliary quarks in
unphysical ``isospin'' doublets and a decoupling of the auxiliary quarks by mass reweighting.}

\FullConference{37th International Symposium on Lattice Field Theory - Lattice2019\\
		16-22 June 2019\\
		Wuhan, China}

\begin{document}
\section{Introduction}

Despite the theoretical and experimental efforts of the past two decades to study the
phase diagram and the properties of QCD at finite quark densities, most of the
parameter space is still uncharted when it comes to first principles results.
The main reason for this is the complex action problem affecting QCD at non-zero baryon
chemical potentials, prohibiting direct simulations of lattice QCD.
Despite the effort to overcome this problem (see~\cite{Gupta:2011ma}, for reviews),
there is currently no reliable method to obtain results close to the pseudo-critical
temperature at physical quark masses.

In the grand canonical ensemble, three-flavor QCD at finite quark density can be described
in terms of baryon ($\mu_B=3[\mu_u+\mu_d]/2$), isospin ($\mu_I=[\mu_u-\mu_d]/2$) and strange
($\mu_s$) chemical potentials. In most of the physical systems all three chemical potentials
are non-zero, while the effects
\begin{wrapfigure}{r}{7cm}
 \centering
 \vspace*{-3mm}
 \includegraphics[width=7cm]{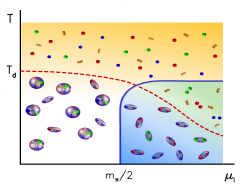}
 \caption{Conjectured phase diagram of QCD at pure isospin
 chemical potential (taken from~\cite{Brandt:2017oyy}).}
 \label{fig:phdiag}
\end{wrapfigure}
due to non-zero $\mu_B$ are usually expected
to dominate. In contrast to QCD at non-zero $\mu_B$ (and $\mu_s$), QCD
at pure isospin chemical potential, i.e., $\mu_B=\mu_s=0$ and $\mu_I\neq0$, with mass
degenerate light quarks has a real and positive fermionic weight factor and, thus,
permits direct simulations. After the first pioneering
studies~\cite{Kogut:2002tm,deForcrand:2007uz,Endrodi:2014lja},
we have recently presented the first results for the continuum phase diagram,
see Fig.~\ref{fig:phdiag}, with physical
quark masses in a realistic three-flavour setup~\cite{Brandt:2017oyy}, showing
Bose-Einstein condensation (BEC) of charged pions at large $\mu_I$ and small
temperatures~\cite{Son:2000xc} (see also Refs.~\cite{Brandt:2018wkp} for further
applications of simulations at non-zero $\mu_I$).

In general, it is the equation of state and the phase diagram at non-zero
$\mu_B,\,\mu_s$ and $\mu_I$ which is of direct phenomenological importance. Since
direct simulations are impossible, one is forced to use indirect methods, such as
the Taylor expansion method and reweighting, to take the effects of non-zero
$\mu_B$ and $\mu_s$ into account. These methods are restricted to
small values of $\mu_B$ and $\mu_s$, rendering the important region beyond nucleon
production threshold inaccessible. In this proceedings article, we apply the
reweighting method to the region around the $\mu_B=\mu_s=0$ axis at
zero temperature with the aim to trace the
BEC phase boundary at non-zero $\mu_B$ and $\mu_s$. In particular, we test two
different reweighting techniques for their efficacy. First, we use a reweighting
in chemical potential (Sec.~\ref{sec:murew}), starting from simulation points
at non-zero $\mu_I$. Secondly, we present a novel technique
(Sec.~\ref{sec:qm-rew}), which employs a reweighting in the
quark mass to decouple auxiliary quarks (forming ``isospin'' doublets with the
physical quarks) from the theory.

\section{Simulations at non-zero $\mu_I$}
\label{sec:sims}

The fermion matrix for a mass-degenerate fermion doublet of staggered quarks $a$ and $b$
at pure isospin chemical potential $\mu_I$ is given by
\begin{equation}
	\label{eq:M}
	\mathcal{M}_{ab} = \begin{pmatrix}
	\slashed{D}_{\mu_I} + m_{ab} & \lambda \eta_5  \\ -\lambda \eta_5 & \slashed{D}_{-\mu_I} + m_{ab}
	\end{pmatrix} ,
\end{equation}
where $\eta_5 = (-1)^{n_t+n_x+n_y+n_z}$ and $\slashed{D}_{\mu}$ is
the discretized massless Dirac operator at chemical potential $\mu$ and including stout smeared links.
In QCD and our standard lattice setup and up to Sec.~\ref{sec:qm-rew}, we
associate $a=u$ and $b=d$ and set $m_{ab}=m_{ud}$ to its physical value. We also include a physical
strange quark at $\mu_s=0$ and use the Symanzik improved gluon action $S_G$.
For $\mu_I\neq0$, the invariance under chiral $\mathrm{SU}(2)_V$ rotations is broken down to a
$\mathrm{U}(1)_{\tau_3}$ subgroup, which is spontaneously broken
for $\mu_I\geq m_\pi/2$ (in our convention for $\mu_I$, see~\cite{Brandt:2017zck}), leading to
the condensation of charged pions.
The unphysical off-diagonal terms in Eq.~(\ref{eq:M}) break the residual $\mathrm{U}(1)_{\tau_3}$
symmetry explicitly.
The explicit breaking serves two main purposes: (i) it enables to observe spontaneous symmetry
breaking in a finite volume; (ii) it acts as a regulator for lattice simulations in the BEC
phase~\cite{Kogut:2002tm,Endrodi:2014lja}. Physical results are obtained
in the limit $\lambda\to0$.

Since simulations are done at $\lambda\neq0$, i.e., including the symmetry breaking terms
in Eq.~(\ref{eq:M}), the full reweighting to the target ensemble includes reweighting factors
$R_\lambda$ for the reweighting to $\lambda=0$. If $O$ is the observable evaluated with target
ensemble parameters, the reweighted result is given by
\begin{equation}
        \label{eq:exp-rew}
	\braket{O}^{\rm targ}_{\lambda=0} = \frac{\braket{O \: R^{\rm targ} R_\lambda}_\lambda}
	{\braket{R^{\rm targ} R_\lambda}_\lambda} \,,
\end{equation}
where $R^{\rm targ}$ is the reweighting factor to target ensemble parameters. In practice, we
will replace the reweighting factor $R_\lambda$ by its leading order expansion in $\lambda$,
as implemented in the improvement program for the $\lambda$-extrapolation~\cite{Brandt:2017zck}.
Consequently, there will remain a residual $\lambda$-dependence for the result on the
left-hand-side of Eq.~(\ref{eq:exp-rew}), which can be removed in terms of an extrapolation.
Note, however, that the leading order reweighting typically captures most of the effects of 
the full reweighting factor, so that the effect of the remaining extrapolation is expected
to be small compared to the effect of the reweighting to the target ensemble. Here we will
focus on the reweighting to the target ensemble and skip the final $\lambda$-extrapolation
for now.

\section{Reweighting in chemical potentials}
\label{sec:murew}

\begin{wrapfigure}{r}{7cm}
 \centering
 \vspace*{-16mm}
	\includegraphics[width=0.47\textwidth]{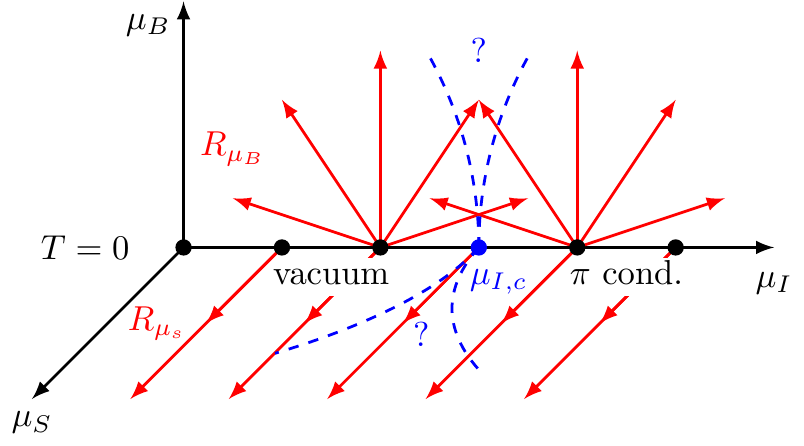}
	\caption{We simulate along the isospin direction and employ reweighting (red) to
	explore the BEC phase boundary (blue) in the vicinity of the $\mu_B=\mu_s=0$ axis.}
	\label{fig:mu_reweighting_idea}
\end{wrapfigure}
We first discuss the reweighting in chemical potentials in the vicinity of the
$\mu_B=\mu_s=0$ axis. In particular, we are interested in the phase diagram and
the behavior of the BEC phase boundary with $\mu_B$ and $\mu_s$. The reweighting
strategy is sketched in Fig.~\ref{fig:mu_reweighting_idea}. The ideal observable to
study the BEC phase boundary would be the pion condensate, but its direct evaluation
in the target ensemble at $\lambda=0$ is difficult. Instead we focus on the
chiral condensate and the isospin density,
\begin{equation}
	\label{eq:rew-obs}
	\bar\psi\psi = \bar{u}u + \bar{d}d \,, \qquad n_I = n_u -n_d \,,
	\qquad \bar{q}q = \frac{T}{V} \frac{\partial \log \mathcal{Z}}{\partial m_{q}}
	\quad \text{and} \quad
	n_q = \frac{T}{V} \frac{\partial \log \mathcal{Z}}{\partial \mu_q} \,.
\end{equation}
At zero temperature, indications for the behavior of the BEC phase boundary can be
obtained from lines of constant observables due to the Silver blaze phenomenon.

\subsection{Reweighting setup}
\label{sec:rew-setup}

The reweighting to the target ensemble is done by reweighting either of the
fermion determinants in the partition function at $\lambda=0$ to a different
value of $\mu_q$, starting from $\mu_u=-\mu_d=\mu_I$
and $\mu_s=0$. To reduce numerical costs, we use the determinant reduction
formalism from~\cite{Toussaint:1989fn, Fodor:2001pe}, where the reweighting
factor from $\mu\to\mu'$ is written as
\begin{equation}
	\label{eq:mu_reweighting}
	R^{\rm targ}(\mu',\mu) = \left(\frac{\det M(\mu')}{\det M(\mu)}\right)^{1/4} =
	e^{-3V_s L_t(\mu'-\mu)/4}\prod_i
	\left(\frac{p_i-e^{L_t\mu'}}{p_i-e^{L_t\mu}}\right)^{1/4} \,,
\end{equation}
where the $p_i$ are the eigenvalues of a reduced matrix $P$, depending on $m_q$,
$V_s$ is the spatial volume and $L_t$ is the temporal extent.
Once the $p_i$ are known, the reweighting in $\mu$ is analytic and can be
done for different values of $\mu'$ at once. In practice, we employ the rooting to each individual term in the
product of Eq.~(\ref{eq:mu_reweighting}), selecting the complex roots which ensure that the determinant is
real and positive at $\mu_B=\mu_s=0$ and continuous along the real $\mu_B$ or $\mu_s$ axis. All results
presented in the following have been obtained on $8^4$ lattices.

\begin{figure}[t]
	\centering
	\includegraphics[width=0.4\textwidth]{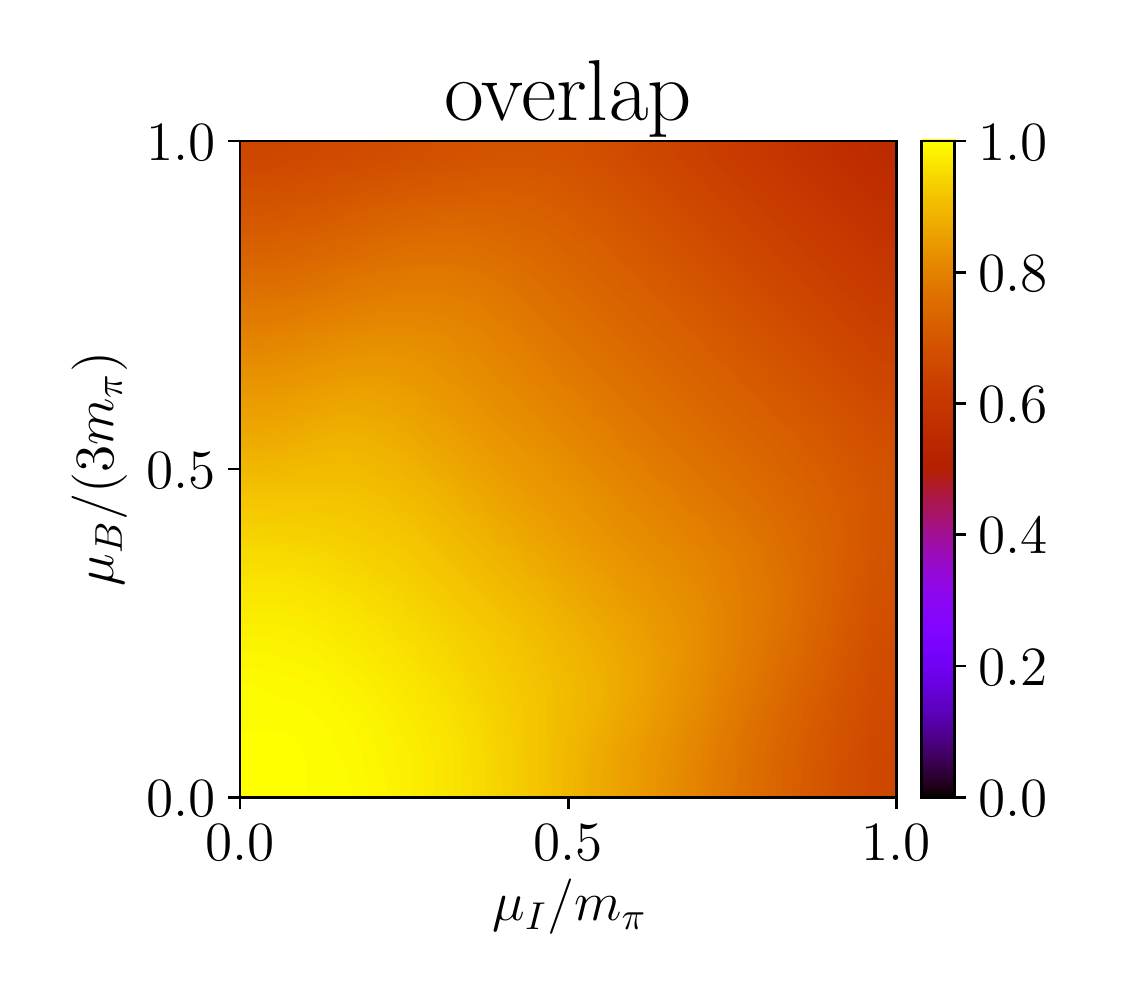}%
	\hfil
	\includegraphics[width=0.4\textwidth]{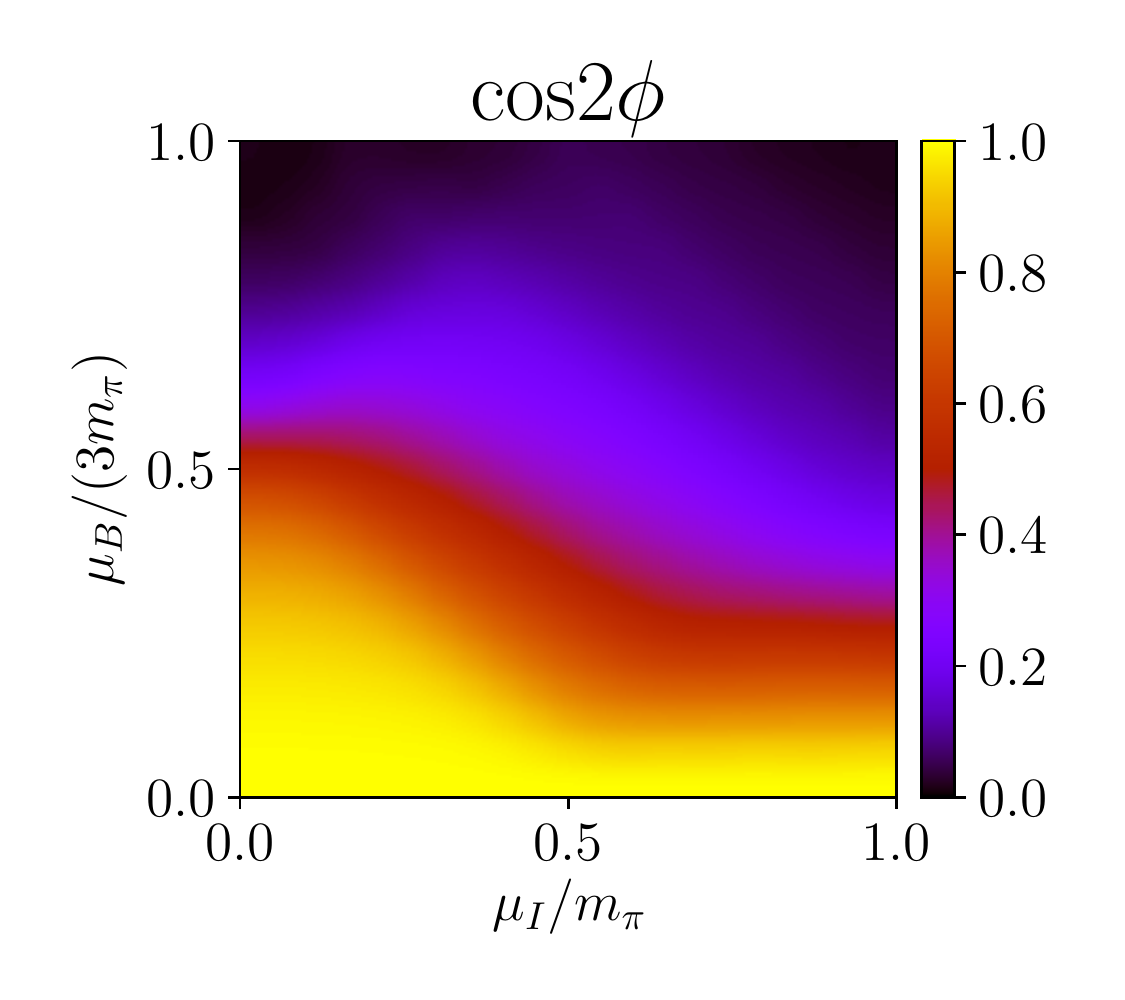}%
	\caption{Overlap (left) and sign problem (right) obtained by reweighting
	from $\mu_B=\mu_I=0$.}
	\label{fig:reweighting_reliability}
\end{figure}

The parameter range to which the reweighting method can be applied is limited by the
overlap between initial and target ensembles and the severity of the sign problem.
To estimate the overlap $\gamma$, we extend the argument
from~\cite{Csikor:2004ik, Schmidt:2004ke} to complex reweighting factors. Considering a
sorted set of reweighting factors $R_i$ ($|R_1| \geq \ldots \geq |R_N|$), which are
normalized, $\sum_i |R_i| = 1$, we define $\gamma$ implicitly via
$\sum_{i=1}^{N\gamma/2} |R_i| = 1-\gamma/2$.
If $\gamma=1$ all configurations have similar weights, i.e., the overlap is large.
If only a few configurations effectively contribute $\gamma$ will be close to zero
and expectation values can be biased. The severeness of the sign problem can be
assessed from the phase fluctuations of the reweighting factor $R= |R|e^{i\phi}$,
$\cos (2\phi) = \text{Re} R^2/|R|^2$. Values close to unity indicate small
phase fluctuations, whereas small values indicate a severe sign problem.
Both quantities are shown for the reweighting from $\mu_B=\mu_I=0$ to $\mu_B,\,\mu_I\neq0$
in Fig.~\ref{fig:reweighting_reliability}. In the following we only present results with
$\gamma>0.5$ and where the partition function is real and positive, another indication
that the sign problem is under control.

\subsection{Results}

\begin{figure}[t]
	\centering
	\includegraphics[width=0.5\textwidth]{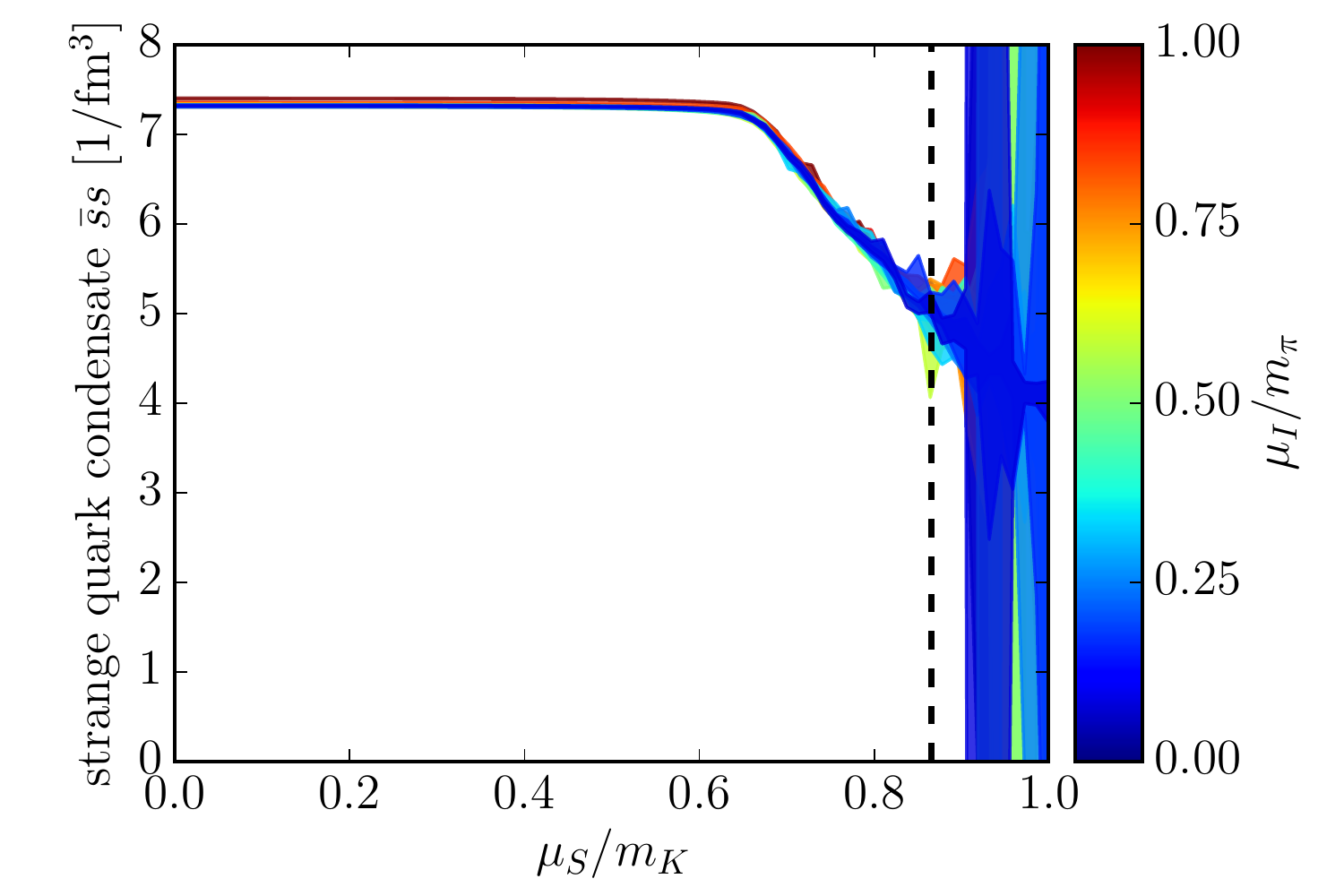}%
	\includegraphics[width=0.5\textwidth]{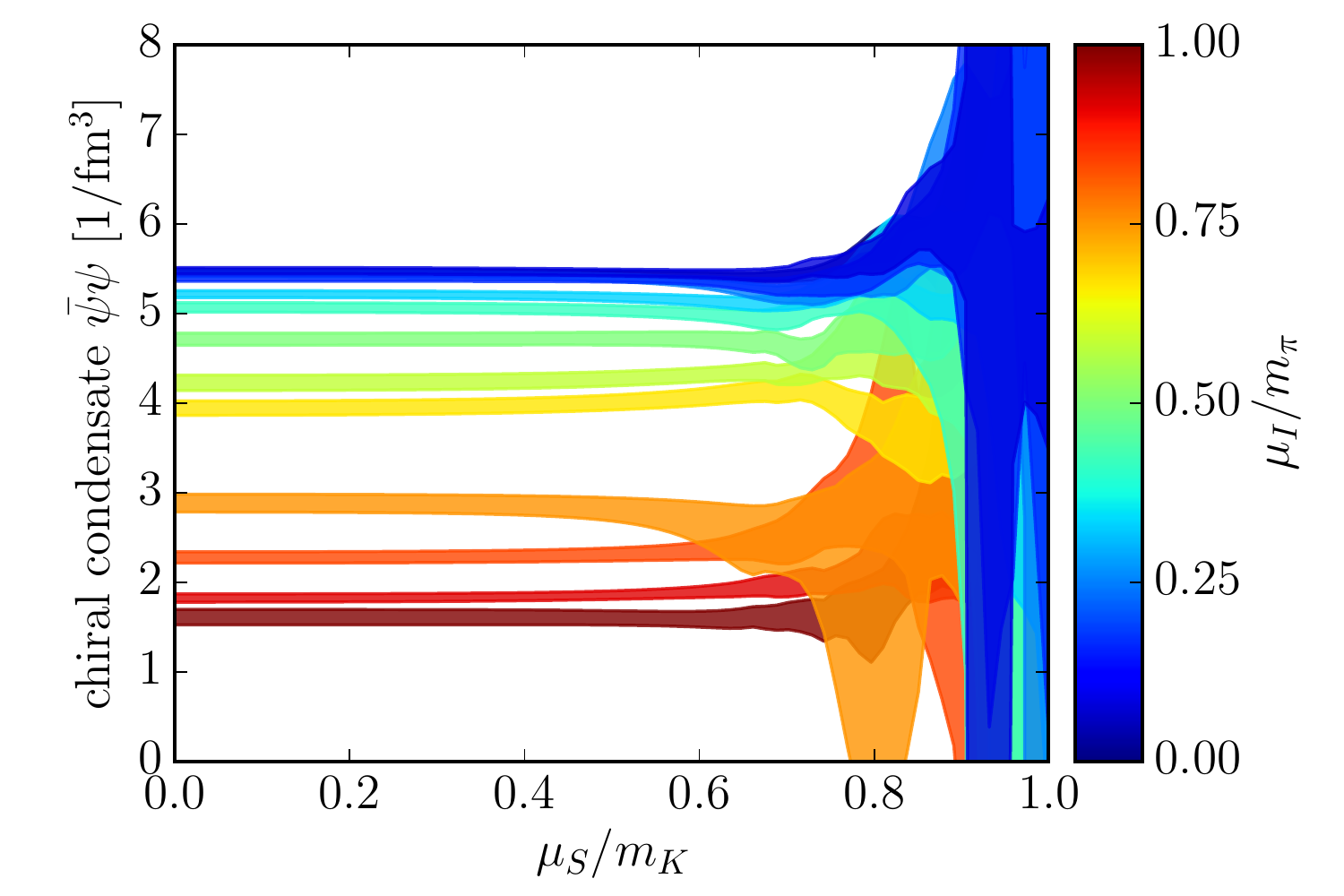}%
	\caption{Effect of $\mu_s$ on the $\braket{\bar{s}s}$ (left) and $\braket{\bar{\psi}\psi}$
	(right) for different values of $\mu_I$ (colour code). The dashed line on the left
	indicates the limit for kaon condensation~\cite{Mammarella:2015pxa}.}
	\label{fig:reweighting_muS}
\end{figure}

We will first discuss the results obtained at non-zero $\mu_s$. The
reweighted results for the strange quark and chiral
condensates are displayed in Fig.~\ref{fig:reweighting_muS}. For all values of $\mu_I$
the strange quark condensate drops significantly for $\mu_s>0.6 \cdot m_K$, which could be an
indication for a phase transition. The chiral condensate is mostly unaffected
by $\mu_s$. In~\cite{Mammarella:2015pxa} it has been proposed that the onset of kaon condensation
depends on $\mu_I$ and does not happen for $\mu_S < 0.865 \cdot m_K$, indicated by the dashed
line in Fig.~\ref{fig:reweighting_muS}. Unfortunately, reweighting looses its efficacy in this region.
It might be that our signal is a precursor effect to kaon condensation due to finite size and
temperature effects, but this will have to be clarified in the future. Below
$\mu_s=0.6 \cdot m_K$ we clearly do not observe any effect of $\mu_s$ on the BEC phase boundary.

\begin{wrapfigure}{r}{7cm}
 \centering
 \vspace*{-6mm}
 \includegraphics[width=7cm]{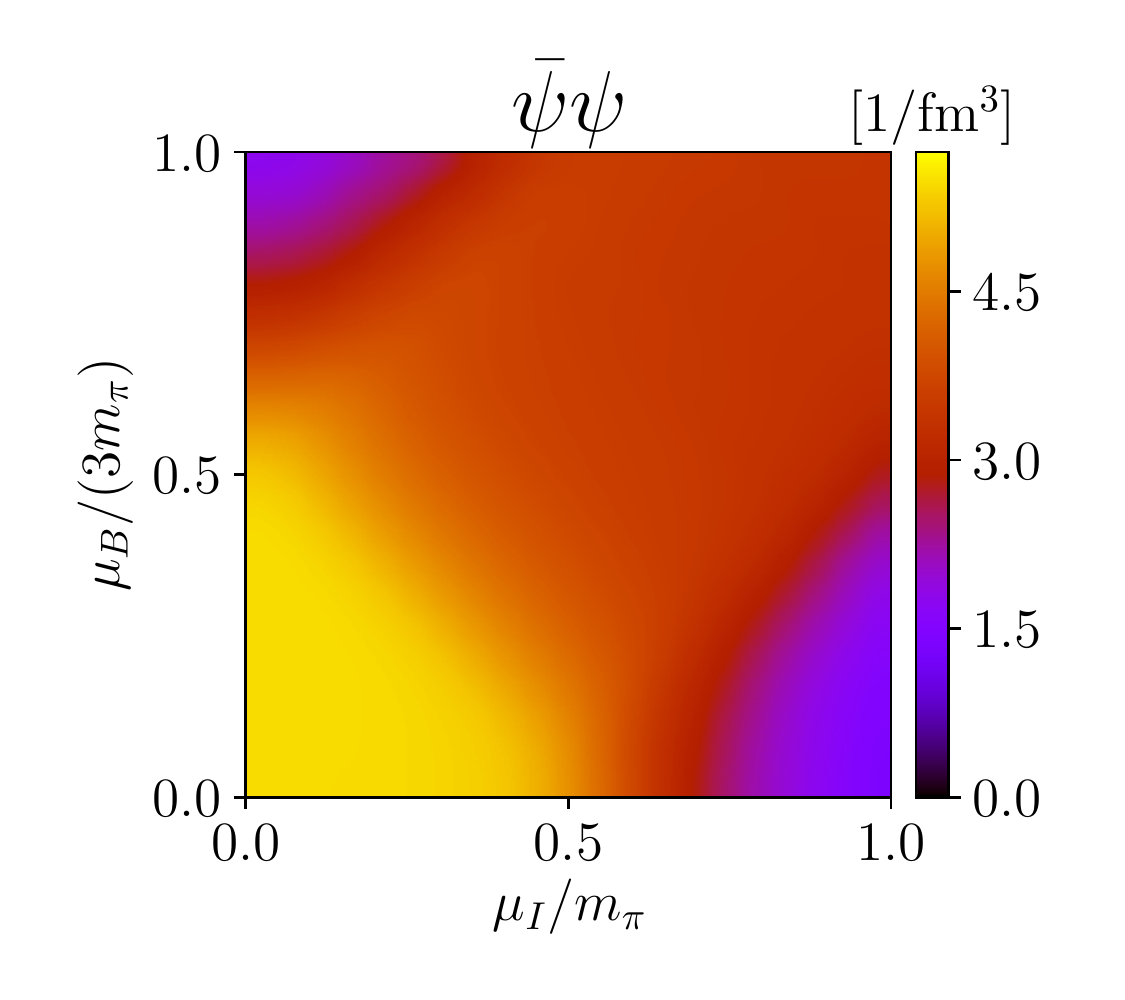}
 \caption{The chiral condensate at non-zero $\mu_B$.
 We combined several auxiliary ensembles in this plot, only considering the ones
 with sufficient overlap with the target ensemble and a reasonably controlled sign problem.}
 \label{fig:reweighting_muB}
\end{wrapfigure}
The results for the chiral condensate reweighted to non-zero $\mu_B$ are shown in
Fig.~\ref{fig:reweighting_muB}. In our setup an indicator for the BEC phase boundary is a
strong decrease of the chiral
condensate. We observe that the surface of strong depletion of the condensate (the purple region)
bends towards larger values of $\mu_I$ in Fig.~\ref{fig:reweighting_muB} and we interpret this
as an indicator for a similar bend of the BEC phase boundary. It is unexpected that the chiral
condensate also shows a slight fall off across the line from $(\mu_B=0,\,\mu_I=m_\pi/2)$ to
$(\mu_B=3m_\pi/2,\,m_I=0)$. Due to the Silver Blaze phenomenon one does not expect a depletion
of the chiral condensate outside of the BEC phase up to $\mu_B \approx m_N$ (the
mass of the nucleon). We are currently investigating the origin of this, likely
unphysical, behavior.

\section{Reweighting in quark masses}
\label{sec:qm-rew}

In addition to reweighting in chemical potentials from finite $\mu_I$, we propose a novel
reweighting direction to access the phase diagram. The phase diagram with $N_f=2+1$ dynamical
flavours can, in principle, be accessed by simulating with a $N_f=2+2+1$ setup, including
light quarks at the desired quark chemical potentials, accompanied by auxiliary ``isospin''
partners with negative
\begin{wrapfigure}{r}{8cm}
 \centering
 \vspace*{-1mm}
 \includegraphics[width=8cm]{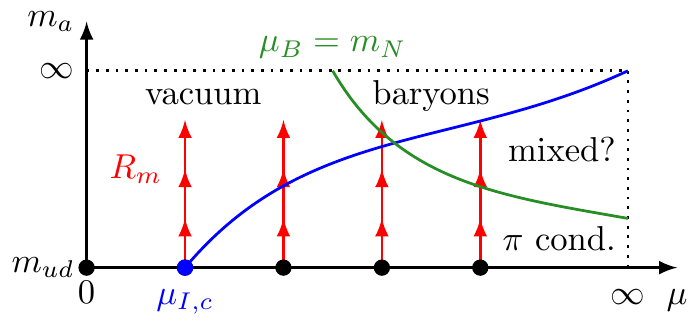}
 \caption{A possible scenario for how the quark masses could affect the phase diagram
	at finite quark chemical potential $\mu$. Changes in the auxiliary quark mass(es)
	influence both the BEC (blue) and baryon creation (green) phase boundaries.
	The red arrows indicate the reweighting direction.}
 \label{fig:mass_reweighting_idea}
\end{wrapfigure}
chemical potential.
The latter are decoupled from the theory by increasing their mass $m_a$ (initially
$m_a=m_{ud}$, for instance) via reweighting and taking the limit $m_a\to\infty$. The underlying
idea and the possible phase diagram for different values of $m_a$ is depicted in
Fig.~\ref{fig:mass_reweighting_idea}.

The reweighting factor for altering one of the quark masses of the quarks in an isospin doublet,
here the one with negative chemical potential, from $m_{ud}$ to $m_a$ is given by 
\begin{equation}
	R^{\rm targ} = \left[ \frac{\det(\Dslash_{-\mu_I} + m_a)}
	{\det(\Dslash_{-\mu_I} + m_{ud})}\right]^{1/4}.
\end{equation}
Since $\Dslash_{-\mu_I}$ is a non-normal operator, the reweighting to different values of $m_a$
can again be done analytically when we know the left and right eigenvectors of $\Dslash_{-\mu_I}$.
The non-normality also needs to be taken into account when computing observables~\cite{Tanaka_2006}.

\begin{figure}[b]
	\centering
	\includegraphics[width=0.5\textwidth]{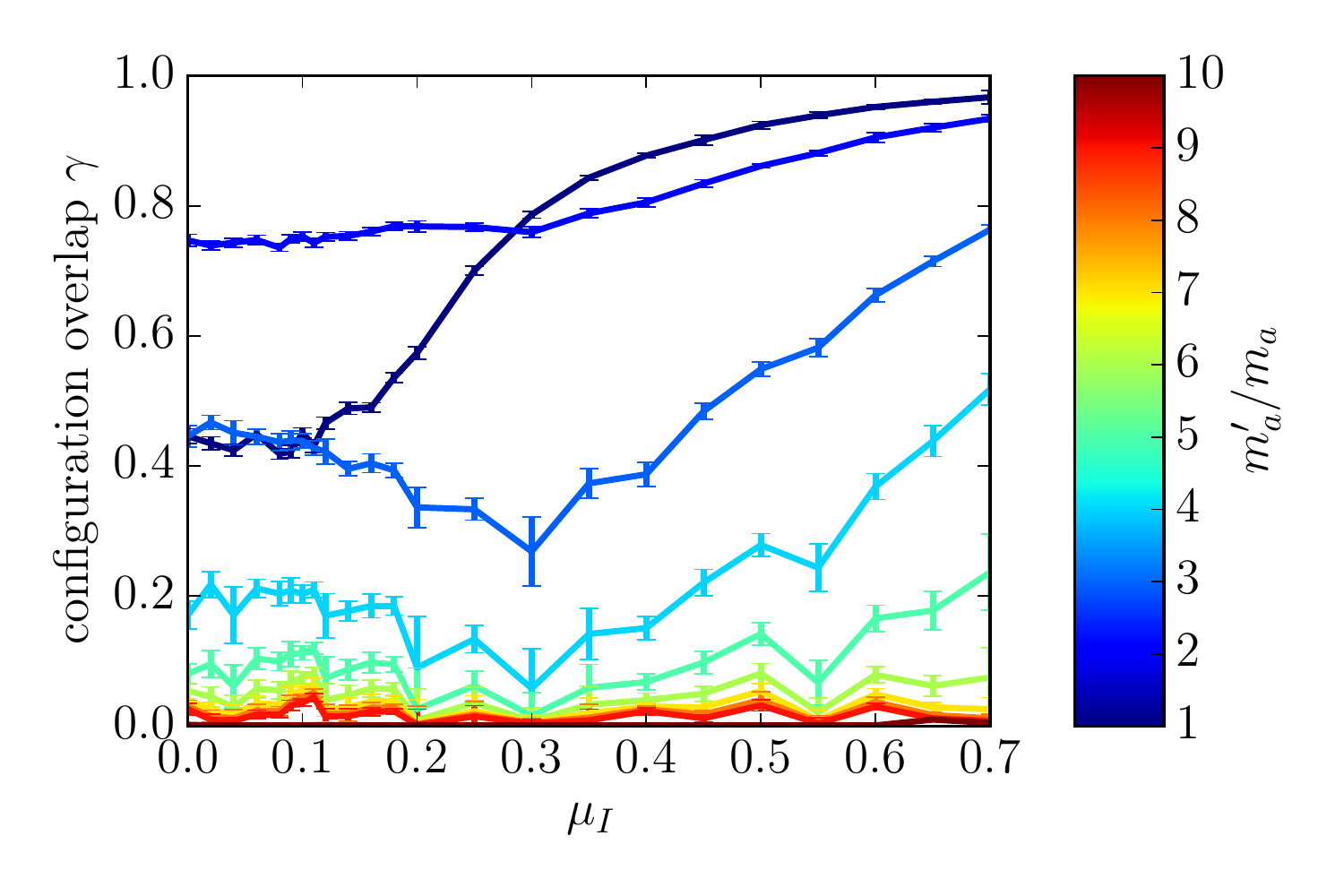}%
	\includegraphics[width=0.5\textwidth]{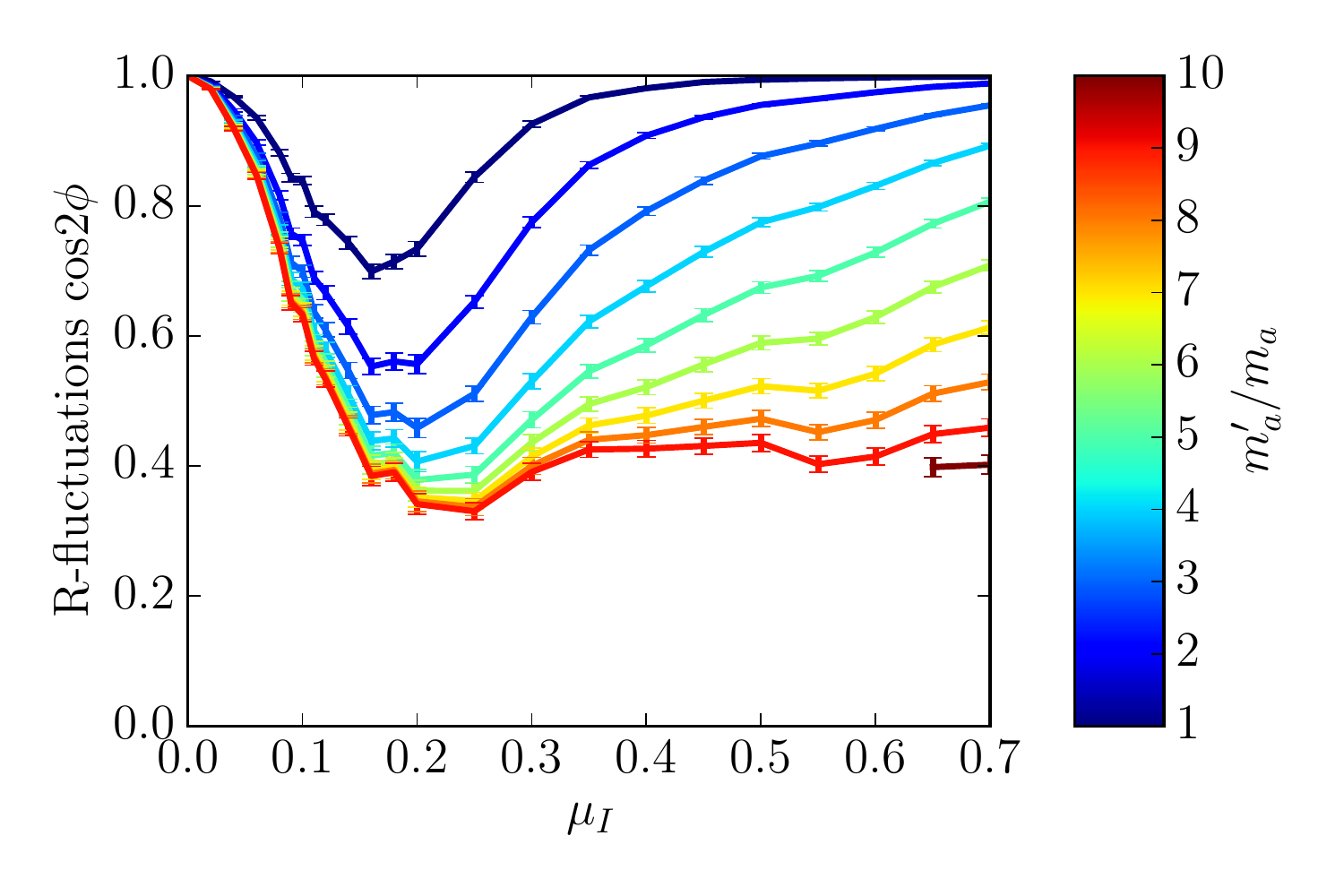}%
	\caption{Overlap (left) and sign problem (right) in case of the mass reweighting described
	in the text.}
	\label{fig:reweighting_mq}
\end{figure}

We observed that adding light quarks induces drastic changes in the system, strongly affecting the lattice spacing,
$m_\pi$ and the value of $\pbp$. To improve the overlap between initial and target ensembles, we tuned
the masses of the isospin doublets in the Silver blaze region such that the values of $\pbp$ roughly
match those of the $N_f=2+1$ ensemble, making them about 11 times heavier. We then reweight the masses
of the two physical quarks (the ones with positive chemical potential) to those of the target ensemble
and decouple the auxiliary quarks by increasing $m_a$. Preliminary results for overlap and sign problem for
such a reweighting are shown in Fig.~\ref{fig:reweighting_mq}. Apparently there is an optimal overlap
between initial and target ensemble
when varying $m_a$ by a factor between one to three. As expected, the sign problem reappears and
becomes stronger with increasing $m_a$.

\section{Summary and Outlook}

In this proceedings article, we presented results for the pion condensation phase boundary
for small baryonic and strange chemical potentials obtained from reweighting in chemical
potentials from ensembles generated at finite isospin chemical potentials. The results are still in
a preliminary stage and have been obtained on small ($8^4$) lattices. We observed an unexpected behavior
of the chiral condensate at non-zero baryon chemical potential and we are currently investigating
possible causes. One issue concerns the ambiguity in evaluating the phase factor of the rooted
reweighting factor of Eq.~(\ref{eq:mu_reweighting}). A new method to avoid the rooting has been
proposed recently~\cite{Giordano:2019gev} and we plan to test it in the future. In addition, we
proposed a novel reweighting in the masses of auxiliary quarks to study QCD at finite quark
chemical potentials and presented first tests of this method.

\end{document}